# Compressed Silica Aerogels for the Study of Superfluid $^3$He

J. Pollanen[a], H. Choi[a], J.P. Davis[a], S. Blinstein[a], T.M. Lippman[a], L.B. Lurio[c],
N. Mulders[b], and W.P. Halperin[a]

[a] *Department of Physics and Astronomy, Northwestern University, Evanston, IL 60208, USA*
[b] *Department of Physics and Astronomy, University of Delaware, Newark, DE 19716, USA*
[c] *Department of Physics, Northern Illinois University, DeKalb, IL 60115, USA*

**Abstract.** We have performed Small Angle X-ray Scattering (SAXS) on uniaxially strained aerogels and measured the strain-induced structural anisotropy. We use a model to connect our SAXS results to anisotropy of the $^3$He quasiparticle mean free path in aerogel.

## INTRODUCTION

Measurements of the low temperature phase diagram of superfluid $^3$He in 98% aerogel indicate a stable *B*-phase and a metastable *A*-like phase [1-3]. Vicente *et al.* proposed that the relative stability of these phases can be attributed to local anisotropic scattering of the $^3$He quasiparticles by the aerogel network [4]. This network consists of silica strands with a diameter of ~30 Å and average separation $\xi_a \approx 300$ Å. Vicente *et al.* also proposed using uniaxial strain of the aerogel to produce global anisotropy [4]. We have performed SAXS on two uniaxially strained aerogels and found that strain introduces anisotropy on the ~100 Å length scale. We relate this to anisotropy of the quasiparticle mean free path, $\lambda$.

## EXPERIMENT

SAXS studies of two aerogel samples (97.1% and 98% porosity) were performed at Sector 8 of the Advanced Photon Source (APS) at Argonne National Laboratory, using a photon energy of 7.5 keV. The 97.1% sample was grown at Northwestern University and the 98% sample was grown at the University of Delaware. The samples were cylinders with diameter to height ratios of 1.53 (97.1%) and 0.65 (98%). For the sample grown at Northwestern radial shrinkage of ~10% was observed after supercritical drying. The porosity of the sample was measured after drying.

The samples were uniaxially strained along the cylinder axis and oriented such that the strain axis was perpendicular to the x-ray beam. For the 98% aerogel, SAXS was performed with nominally zero and 28% strain. For the 97.1% sample, the sample conditions were fully relaxed and a series of increasing strains in the range 3.5-52.8%. Beyond ~30% the sample showed significant damage in the form of cracks and the results were not analyzed. For each value of strain the scattered x-ray intensity, $I(q)$, was obtained, where $q$ is the scattered x-ray wave vector. $I(q)$ was binned for various values of the azimuthal angle, $\phi$, defined with respect to the incident x-ray beam in the plane of the CCD camera. $\phi = 90°$ is parallel to the strain axis.

## RESULTS AND DISCUSSION

Our analysis of the scattering curves is based on a phenomenological scattering function used to fit $I(q)$,

$$I(q) = \frac{C\xi^d}{\left(1+q^4\xi^4\right)^{d/4}} \frac{\left(1+q^2\xi^2\right)^{1/2}}{q\xi} \quad (1)$$
$$\times \sin\left[(d-1)\tan^{-1}(q\xi)\right],$$

where $C$, $d$, and $\xi$ are fit parameters. Eq. (1) is a modified version of the structure factor described by Freltoft *et al.* for a fractal structure [5]. In Eq. (1), $C$ is a constant, $d$ is approximately the fractal dimension of the aerogel, and $\xi$ is associated with the upper length scale at which the aerogel ceases to be fractal in nature. This length scale is of the order of $\xi_a$. In the limit $q\xi \gg 1$, Eq. (1) is proportional to $q^{-d}$. This procedure produced fits that match the data well. Fig. 1 presents $I(q)$ for a strain of 21.1% ($\phi = 90°$, $\phi = 174°$) along with fits produced by Eq. (1).

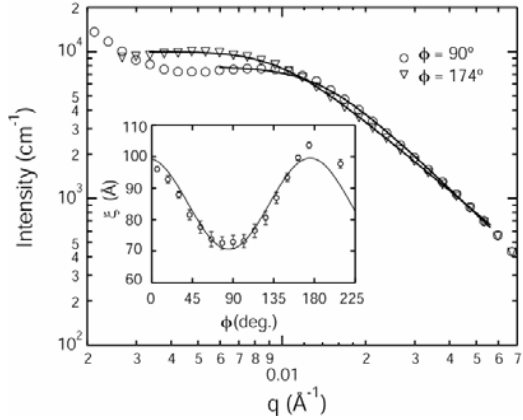

**FIGURE 1.** $I(q)$ fit with Eq. (1) ($\phi = 90°$, 174°) for the 97.1% aerogel strained by 21.1%. The inset depicts $\xi(\phi)$ for this strain.

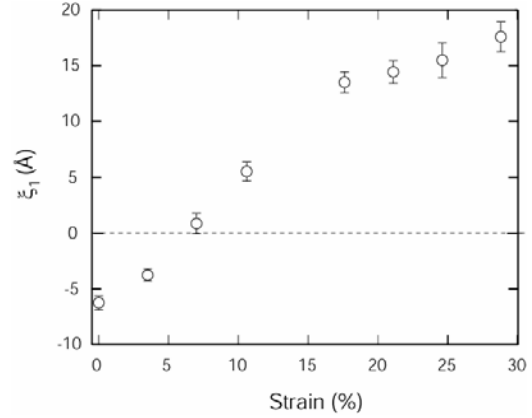

**FIGURE 2.** $\xi_1$ vs. strain for the 97.1% aerogel.

For each value of strain we plotted $\xi$ vs. $\phi$, and found it to vary as $\xi(\phi) = \xi_0 + \xi_1 \sin(2\phi + \theta)$. The results for the 97.1% sample strained by 21.1% are presented in the inset of Fig. 1. For this sample we found $\theta = 99.4° \pm 1.3°$ close to the expected symmetry axis of $\theta = 90°$. $\xi_0$ ranged from 85-98 Å. Similar $\phi$-dependence was found for $C$ and $d$. A consistent sinusoidal behavior was observed for the 98% porosity aerogel compressed by 28%. We found that $\theta = 96.4° \pm 1.5°$ for this sample. Note, the error estimates on $\theta$ are statistical. Previous SAXS studies on isostatically strained aerogels also indicate a decrease in $\xi$ with strain [6]. The increase of $\xi_1$ with strain, shown in Fig. 2, demonstrates that anisotropy can be introduced into the aerogel.

For the 97.1% sample we discovered intrinsic anisotropy present with the sample unstrained. $\xi(\phi)$ was ~180° out of phase relative to the strained case. We suspect that this intrinsic anisotropy was produced during the synthesis of the aerogel. Slight anisotropy in the nominally unstrained 98% sample was observed but might be accounted for by the strain from the sample holder. For the unstrained 97% sample there was no stress from the sample holder.

We have used a simple model of the aerogel network to connect the anisotropy measured in $\xi$ with the anisotropy of the quasiparticle mean free path [7]. We find, $\lambda_\perp/\lambda_\parallel = 2(\xi_\perp/\xi_\parallel)/((\xi_\perp/\xi_\parallel)+1)$, where $\lambda_\perp(\lambda_\parallel)$ is the mean free path perpendicular (parallel) to the strain axis and $\xi_\perp \equiv \xi_0 + \xi_1$ and $\xi_\parallel \equiv \xi_0 - \xi_1$. At a strain of 28% $\lambda_\perp/\lambda_\parallel = 1.33$ for the 98% sample and $\lambda_\perp/\lambda_\parallel = 1.21$ for the 97.1% sample. In conclusion, uniaxial strain produces significant global anisotropy in the structure of the aerogel and may be used to test the relative stability of the $A$ and $B$ phases of superfluid $^3$He. Preliminary measurements to this effect by J.P. Davis et al. have been conducted in 98% aerogel grown at Northwestern [8].

## ACKNOWLEDGMENTS

The authors would like to thank J.A. Sauls for valuable theoretical insights. We are grateful to G.W. Scherer and J.F. Poco for their advice regarding aerogel shrinkage and fabrication. We also acknowledge the support of the National Science Foundation, DMR-0244099.